\documentclass[12pt]{article}
\usepackage[body={17cm, 23cm, centered}]{geometry}

\usepackage{lmodern}
\usepackage{epsfig,verbatim,cite}
\usepackage{braket}

\usepackage[T1]{fontenc}
\usepackage[utf8]{inputenc}
\usepackage[unicode]{hyperref}
\hypersetup{colorlinks=true,linktocpage,breaklinks,
            urlcolor=blue,
            linkcolor=red,
            citecolor=blue
            }

\usepackage{float} 
\usepackage{subcaption}

\usepackage{mathtext,marvosym,textcomp}
\usepackage{mathtools}
\usepackage{slashed}
\usepackage[dvipsnames,svgnames,table]{xcolor}
\usepackage{tikz}
\usepackage[ normalem]{ulem}
\usepackage{tocloft}
\usepackage{epsfig,amsmath,amsfonts,amssymb,arydshln}
\usepackage[makeroom]{cancel}
\usepackage{multirow}
\usepackage{subcaption}

\def\a{\alpha} \def\b{\beta} \def\g{\gamma} \def\d{\delta} \def\e{\epsilon}
  \def\h{\eta} \def\q{\theta}
  \def\k{\kappa} \def\l{\lambda} \def\m{\mu}
 \def\x{\xi} \def\p{\pi}  \def\r{\rho}
 \def\s{\sigma} \def\t{\tau}  \def\f{\phi}
  \def\y{\psi} \def\w{\omega}

   \def\W{\Omega}

\def\fr{\frac}  \def\dt{\partial}

\def\ph{\phantom}
\def\mc{\mathcal}

\def\mH{\mathcal{H}}

\def\RR{\mathbb{R}}

\def\SS{\mathbb{S}}

\def\rmU{\mathrm{U}}

\def\rmO{\mathrm{O}}
\def\rmSO{\mathrm{SO}}

\def\AdS{\mathrm{AdS}}
\def\CP{\mathbb{C}\mathrm{P}}

\newcommand\bqa {\begin{eqnarray}}
\newcommand\eqa {\end{eqnarray}}

\newcommand{\bear}{\begin{array}}
\newcommand{\enar}{\end{array}}

\def\ab{{\bar{a}}}

\newcommand{\be}{\begin{equation}}
\newcommand{\ee}{\end{equation}}
\newcommand{\bea}{\begin{eqnarray}}
\newcommand{\eea}{\end{eqnarray}}

\def\const{\mathrm{const}}

\begin{document}

\numberwithin{equation}{section}

\renewcommand{\contentsname}{}
\renewcommand{\refname}{\begin{center}References\end{center}}
\renewcommand{\abstractname}{\begin{center}\footnotesize{\bf Abstract}\end{center}} 

\begin{titlepage}
\ph{preprint}

\vfill

\begin{center}
   \baselineskip=16pt
   {\large \bf On integrability of tri-vector deformed Type II string
   }
   \vskip 2cm
    Edvard T. Musaev$^{a,b}$\footnote{\tt musaev.et@phystech.edu},
    Gleb S. Zverev$^{a,c}$\footnote{\tt zverev.gs@mipt.ru }
       \vskip .6cm
             \begin{small}
                          {\it
                          $^a$Moscow Institute of Physics and Technology, 
                         141702, Dolgoprudny, Russia\\ 
                          $^b$Kazan Federal University, Institute of Physics, 420111, Kazan, Russia\\
                          $^c$National Research Nuclear University MEPhI, Moscow, 115409, Russia
                          } \\ 
\end{small}
\end{center}

\vfill 
\begin{center} 
\textbf{Abstract}
\end{center} 
\begin{quote}
We analyse dynamics of the closed Type IIA and IIB string on various tri-vector deformed background searching for signatures of integrability. Using numerical methods we construct Poincar\'e sections for particular embeddings of the string and show that these are not destroyed under tri-vector deformation. We find that the corresponding Lyapunov exponents decay showing that trajectories do not diverge. 
\end{quote}

\vfill
\setcounter{footnote}{0}
\end{titlepage}

\tableofcontents
\setcounter{page}{1}

\section{Introduction}

Integrability stands among the most valuable properties of a physical system. For a classical mechanical system integrability implies existence of a set of conserved quantities that allow to rewrite equations of the system in the so-called action-angle variables and integrate them completely. The situation is more tricky for field theories, where one formally needs an infinite (continuous) family of conserved quantities to ''match'' the total amount of degrees of freedom. A consistent treatment of such systems is given by the Zakharov--Shabat formalism stating that if classical equations of motion of a 2d system can be written as the flatness condition of a (Lax) connection, then the system is integrable in the Liouville sense. For theories with periodic boundary conditions the infinite tower of conserved charges is constructed of components of the Lax connection (see e.g. \cite{Torrielli:2016ufi} for a more detailed review). Although here we will be only concerned with classical integrability let us mention that integrability of a quantum system is mostly understood as the possibility to derive the exact energy spectrum. For theories with well defined asymptotically free states integrability means that the full S-matrix solves quantum Yang-Baxter equation, whose quasi-classical limit can be used to construct the corresponding Lax connection. 

Importance of integrability of the two-dimensional string sigma-model is evident in the context of gauge-gravity duality, where it implies integrability of the dual field theory (if the correspondence is taken in the strong sense). Both quantum and classical integrability of the string on various backgrounds have been intensively studied during the last decades and a wide set of fascinating results is currently available. Most relevant to the present paper are the results of \cite{Bena:2003wd}, where integrability of the Type IIB string on the AdS$_5\times \SS^5$ has been shown, and of \cite{Delduc:2013fga,Delduc:2013qra} where a class of integrable deformations of this sigma-model has been presented. This class is given by a family of Yang--Baxter deformations of the initial sigma-model that has been introduced in the context of integrability preserving deformations in \cite{Klimcik:2002zj}. It is worth mentioning that the family of integrable sigma-models found by Cherednik in \cite{Cherednik:1981df} are also in the same family. More detailed discussion of these results and of the recent progress can be found in the reviews \cite{Demulder:2023bux,Seibold:2024qkh}, various approaches to integrability in the context of AdS/CFT correspondence have been extensively reviewed in \cite{Beisert:2010jr}.

Yang--Baxter deformations of a 2d sigma-model can be understood as a field transformation of the underlying supergravity background. Background fields for the ($\h$-)deformed sigma-model of \cite{Delduc:2013fga,Delduc:2013qra} have been found in \cite{Arutyunov:2015qva}, which however appeared to be a solution to a set of equations dubbed generalized supergravity \cite{Hoare:2015wia}. Such understood supergravity side of the picture collocates with the older story of finding supergravity description of exactly marginal deformations, and in particular with the result of \cite{Lunin:2005jy} where a general rule for a(n abelian) bi-vector deformation of the NS-NS fields has been presented:
\begin{equation}
    \label{eq:2transf}
    (g+b)^{-1} = G+\b.
\end{equation}
Here $G$ stands for the initial background metric, $g,b$ are the resulting deformed metric and the b-field, and $\b$ is a bi-vector deformation parameter taken in \cite{Lunin:2005jy} to be $\b = \dt_{\f_1}\wedge \dt_{\f_2}$, i.e. along two commuting isometries of the 5-sphere. As it has been shown in the series of papers \cite{Araujo:2017enj,Araujo:2017jkb,Bakhmatov:2017joy,Bakhmatov:2018apn} more generally one can take $\b = r^{ab}k_a \wedge k_b$ with $r^{ab}$ being a constant matrix and with Killings vectors $k_{a}$ satisfying $[k_a,k_b] = f_{a b}{}^c k_c$. The transformation \eqref{eq:2transf} generates a solution if
\begin{equation}
    \begin{aligned}
        r^{[b_1 [a_1}r^{|b_2|a_2}f_{b_1 b_2}{}^{a_3]} & = 0 ,\\
        r^{ab}f_{ab}^{c}&=0.
    \end{aligned}
\end{equation}
The first line is the classical Yang--Baxter equation, and the second, referred to as the unimodularity condition, has been introduced in \cite{Borsato:2016ose} and is actually the standard unimodularity condition of the dual algebra within the corresponding Drinfeld double. 

Bi-vector Yang--Baxter deformations do not mix NS-NS and R-R fields, that is a consequence of the fact that these are transformation of the O(10,10) group governing T-duality of the 2d sigma-model \cite{Catal-Ozer:2019tmm}. In this paper we are interested in a generalization of bi-vector deformations to U-duality symmetries mixing the NS-NS and R-R sectors known as polyvector deformations \cite{Bakhmatov:2019dow,Bakhmatov:2020kul,Gubarev:2020ydf,Barakin:2024rnz}. For Type IIA theory most straightforwardly these are formulated in terms of the field content of the 11D supergravity, or more precisely in terms of its U-duality symmetric formulation, that is a part of exceptional field theory of \cite{Hohm:2013pua,Hohm:2013vpa,Musaev:2015ces} (for a review see \cite{Baguet:2015xha,Berman:2020tqn,Musaev:2019zcr}). As it is explicitly shown in \cite{Hohm:2013vpa} it appears that exceptional field theory also contains Type IIB theory, allowing to reproduce deformations of the corresponding backgrounds \cite{Gubarev:2024tks}. In 11 dimensions and in Type IIA theory the simplest set of such deformations is parametrized by a tri-vector $\W = \r^{abc}k_a\wedge k_b \wedge k_c$, where the constants $\r^{abc}$ must satisfy a generalization of the classical Yang--Baxter equation
\begin{equation}
\label{eq:gCYBE}
    \begin{aligned}
       \rho^{a_1[a_2|a_6|}\rho^{a_3a_4|a_5|}f_{a_5a_6}^{\ \ \ \, |a_7]}-\rho^{a_2[a_1|a_6|}\rho^{a_3a_4|a_5|}f_{a_5a_6}{}^{|a_7]}=0..
    \end{aligned}
\end{equation}
Similarly to the bi-vector case one usually requires $\r^{abc}f_{ab}{}^d = 0$, that is referred to as a generalized unimodularity constraint. This however is not always necessary as one may generate solutions to ordinary supergravity equations by non-unimodular deformations.

In this paper we address the question of whether tri-vector deformations preserve integrability of the 2d sigma-model. Although the tri-vector generalization of the classical Yang--Baxter equation is structurally very similar to the standard one it is not clear what is its relation to integrability of sigma models (probably, three dimensional). Attempts to sketch possible directions where further understanding can be gained has been presented in the review \cite{Gubarev:2023jtp}. Here we present signatures that 2d sigma model on tri-vector deformed background might be integrable by investigating 
invariant tori in the phase space of the classical string embedded in the deformed background in a particular way, and sign of the corresponding Lyapunov exponent. Such tori are a set of closed phase curves reflecting periodicity of the system and Lyapunov exponent reflects how closed trajectories diverge with time. This approach to our knowledge has been first used in this context in \cite{PandoZayas:2010xpn} to observe signatures of (classically) chaotic behaviour of the closed string on the Schwarzschild--AdS$_5$ background. In \cite{Giataganas:2013dha} the same analysis has been used to show that dynamics of the string on the $\beta$-deformed Lunin--Maldacena with complex deformation parameter $\beta$ exhibits chaos signatures. This is in contrast to deformations with real $\beta$ where no chaos presents (classically) and moreover explicit Lax pair is known. Recently in \cite{Pal:2023bjz} the same approach has been applied to $\gamma$-deformed Gaiotto--Maldacena backgrounds \cite{Gaiotto:2009gz,Nunez:2019gbg}, dual to marginal deformations of $\mathcal{N}=2$ SCFTs, to show that integrability is lost for large enough values of the deformation parameter. Classical dynamics of the Type IIA string on AdS$_4\times \mathbb{C}P^3$ deformed by a bi-vector both along the AdS and compact direction has been considered in  \cite{Pal:2022hmh} where it has been shown that the corresponding invariant tori are stable under the deformation. Here we present further analysis of this system for stability of the tori against particular tri-vector deformations: the abelian $\mathrm{U}(1)^3$ deformation constructed in \cite{Lunin:2005jy} and the non-abelian $PPM$ deformation constructed in \cite{Bakhmatov:2020kul}.

The paper is structured as follows. In Section \ref{sec:basics} we give some details on the Hamiltonian formalism for the string, that we are using, and consider the basic example of the LM deformation of the $\AdS_5\times \SS^5$ background. Main result are presented in Section \ref{sec:trivec} where abelian and non-abelian tri-vector deformation of $\AdS_4\times \CP^3$ are considered. In this section we provide plot of the corresponding Poincar\'e section, Lyapunov exponents and provide analytical results where possible. The Section \ref{sec:conclusions} contains concluding remarks, discussion of the results and of further directions.

\section{2d \texorpdfstring{$\sigma$}{sigma}-model and invariant tori}
\label{sec:basics}

\subsection{The general set-up }

In this section we illustrate the approach by considering the well known classically integrable model of the Lunin--Maldacena deformation of the string on the $\AdS_5 \times \SS^5$ background. Lax pair for this system has been constructed in \cite{Frolov:2005dj} in the Metsaev--Tseytlin formalism, that takes into account full dynamics on the supercoset manifold $\fr{SU(2,2|4)}{SO(1,4)\times SO(5)}$. In the numerical approach utilized below one is certainly not able to consider the full string dynamics in its generality and has to specify an embedding into the target space. Therefore for our purposes it is sufficient to consider only the bosonic part of the string action, meaning that we keep fermionic degrees of freedom unexcited. Hence, we are working with the  classical 2d sigma-model in the conformal gauge on a general 10d background given by the metric $g_{MN}$ and the Kalb--Ramond field $B_{MN}$:
\begin{equation}
    S_P = -\fr12 \int d\sigma d\tau \Big(\sqrt{-h} h^{\a\b}g_{MN} + \e^{\a\b}b_{MN}\Big)\dt_\a X^M \dt_\b X^N,
\end{equation}
where $\a,\b = 0,1$ are world-sheet indices, $h^{\ab}$ is the inverse world-sheet metric, $\e^{01}=-1$ and capital Latin indices $M,N=0,\dots,9$ label directions in the target space-time. Canonical momentum is defined in the usual way and reads
\begin{equation}
    P_M = \fr{\d S}{\d \dt_0 X^M} = - \big(\sqrt{-h}h^{0\a} g_{MN} + \e^{0\a}b_{MN}\big)\dt_\a X^N.
\end{equation}
The system possesses gauge symmetries and its dynamics is restricted by a set of two primary constraints. These can be derived by varying the action with respect to the world-sheet metric $h^{\a\b}$
\begin{equation}
    T_{\a\b} = \fr{\d S}{\d h^{\a\b}} = 0.
\end{equation}
Of the three components only $T^0{}_0$ and $T^1{}_0$ are functionally independent and are used as the constraints:
\begin{equation}
    \begin{aligned}
        C_1 & = P_M \dt_\s X^M, \\
        C_2 & = g^{MN}\big(P_M - b_{MK}\dt_\s X^K\big)\big(  P_N - b_{NL}\dt_\s X^L\big) + g_{MN}\dt_\s X^M \dt_\s X^N.
    \end{aligned}
\end{equation}
The canonical Hamiltonian derived from the action $S_P$ is identically zero since it is proportional to the constraints. To see that it is suggestive to rewrite the action factoring out the combinations $C_1$ and $C_2$. To do so one expresses $\dt_0 X^M$ in terms of the canonical momentum $P_M$ and adds and subtracts the term $P_M \dot{X}^M$ where the dot denotes derivative w.r.t. to $\t = \s^0$. This gives the following expression
\begin{equation}
    S_P = \int d^2\s \Big(P_M \dot{X}^M + \fr{1}{\sqrt{-h}h^{00}}C_2 + \fr{2 h^{01}}{h^{00}}C_1\Big).
\end{equation}
When defining dynamics of such a system with constraints one adds Lagrange multipliers, investigates algebra of the primary and secondary (if any) constraints and defines Dirac bracket. Since we aim at explicit solutions to classical equations of motion rather than a general analysis, we instead  gauge fix the action and derive the corresponding Hamiltonian function. This will define dynamics of the remaining degrees of freedom.

Let us first turn to light cone coordinates $X^M=(X^+,X^-,X^m)$ and assume $g_{\pm m}=0$, $g_{MN} = g_{MN}(X^+,X^m)$ and $b_{+-}=0$, $b_{\pm  m} =0$ that is always true for our examples\footnote{The Type IIB example does not fit this requirement, however its particular field configuration is such, that the same simplification hold and the resulting Hamiltonian is the same up to a numerical factor.}.  We use the world-sheet diffeomorphism and Weyl invariance  to bring the world-sheet metric to the form $h_{\a\b} = \mathrm{diag}[-1,1]$ and the isometry along $X^-$ to fix the light-cone gauge
\begin{equation}
    X^- = \tau.
\end{equation}
This fixes the momentum  $P_+ = g_{+-}$, and for the constraints we have
\begin{equation}
    \begin{aligned}
        C_1 & = P_+ \dt_\s X^+ + P_m \dt_\s X^m,\\
        C_2 & = g^{+-}P_+ P_-  + \mH_x,
    \end{aligned}
\end{equation}
where 
\begin{equation}
    \mH_x = g^{mn}\big(P_m - b_{mk}\dt_\s X^k\big)\big(  P_n - b_{nl}\dt_\s X^l\big) + g_{mn}\dt_\s X^m \dt_\s X^n.
\end{equation}
Dependence of $X^+$ on $\s$ is fixed by the constraint $C_1 = 0$ while its dependence on $\tau$ gives the momentum
\begin{equation}
    P_- = g_{+-}\dot{X}^+.
\end{equation}
Substituting this into the action we have 
\begin{equation}
    \begin{aligned}
    S_P & = \int d^2\s \big(P_+\dot{X}^+ + P_- + P_m \dot{X}^m\big) \\
     & \int d^2\s \big(2 P_- + P_m \dot{X}^m\big),
    \end{aligned}
\end{equation}
reading that the gauge fixed Hamiltonian is $\mH = -2 P_-$ and dynamical degrees of freedom are $X^m$. Finally, using the constraint $C_2 = 0$ we obtain for the Hamiltonian density
\begin{equation}
    \mH = 2\mH_x = 2g^{mn}\big(P_m - b_{mk}\dt_\s X^k\big)\big(  P_n - b_{nl}\dt_\s X^l\big) + 2g_{mn}\dt_\s X^m \dt_\s X^n.
\end{equation}
Hamiltonian of the system is then $H=\int d\s \mH $ and equations of motion are given in terms of Poisson brackets. For the canonical variables the only non-vanishing brackets read
\begin{equation}
    \{P_m(\s),X^n(\s')\} = \d_m{}^n \d(\s-\s').
\end{equation}

A particular embedding will be specified for a given target space further, however the general idea is to get rid of oscillator excitations on the string and to wrap it on cycles of the target manifold. In this form Hamiltonian dynamics of the closed string becomes very similar to that of a point particle and hence the standard formalism of invariant tori is applicable. The idea behind the formalism is simple: evolution of an integrable system with given initial conditions is described by a trajectory that wraps an invariant torus in the phase space \cite{Arnold:1989who,Kolmogorov:1958kol}. For a one dimensional mechanical system one is able to draw full phase trajectories that will be represented by almost closed curves. For systems with more that one canonical variables this is no longer possible, and instead one investigates the so-called Poincar\'e section. These are two-dimensional surfaces in the full phase space (planes in our case) whose intersection with (a set of) phase curves gives its section of an invariant torus. Therefore, for an integrable system one expects the corresponding Poincar\'e section to have a regular structure of closed  curves. 

Another signature of integrability that we investigate is the so-called Lyapunov exponent, that measures distance between two phase trajectories starting at points with slightly different initial conditions. If the distance indefinitely grows with time, the system tends to exhibit chaotic behavior, on the contrary if the exponent decreases with time it is likely that the system is integrable. In other words, decreasing Lyapunov exponent indicates that small variation in initial conditions evolves into a small difference between final state points in the phase space.

It is important to note here, specifying an particular embedding we are investigating integrability of such a subsector of the full theory, that can be identified with a theory of a point-like object. On principle, integrability of any such a truncation does not imply integrability of the full theory in any sense. Actually, the opposite is also true: one may loose integrability of a system by truncating its Hamiltonian. The classical example would be equations of the Toda system whose integrability is lost if the potential is truncated to any finite number of terms. However, stability of invariant tori under a deformation cannot be seen merely as a coincidence, as for example complex $\b$-deformations analysed in \cite{Giataganas:2013dha} do not demonstrate such a property. Therefore, our results for tri-vector deformations presented below should be understood as demonstrating signatures of integrability (in the Liouville sense) of the 2d sigma-model on the corresponding backgrounds, justifying further efforts in searches of the Lax connection and/or analysis of scattering amplitudes. We comment on this more in the concluding section.

\subsection{A transparent example: integrable deformation of \texorpdfstring{$\mathrm{AdS}_{5}\times\mathbb{S}^{5}$}{AdS5xS5}}

To demonstrate the procedure, let us consider a transparent and well understood example of an integrable deformation of a solution of 10-dimensional supergravity, namely the Lunin--Maldacena deformation of the $\mathrm{AdS}_{5}\times\mathbb{S}^{5}$ background  \cite{Lunin:2005jy}. The deformed background is given by 
\begin{equation}
    \begin{aligned}
    ds^{2}_{10} & = R^{2}\left(ds^{2}_{\mathrm{AdS}_{5}} + \sum_{i=1}^{3}\left(d\mu^{2}_{i} + G\mu^{2}_{i}d\phi^{2}_{i}\right) + \hat{\gamma}^{2}G\mu^{2}_{1}\mu^{2}_{2}\mu^{2}_{3}\left(\sum_{i=1}^{3} d\phi_{i}\right)^{2}\right)   \\
    B & = \hat{\gamma}R^{2}G\left(\mu^{2}_{1}\mu^{2}_{2}d\phi_{1}\wedge d\phi_{2} + \mu^{2}_{2}\mu^{2}_{3}d\phi_{2}\wedge d\phi_{3} + \mu^{2}_{3}\mu^{2}_{1}d\phi_{3}\wedge d\phi_{1}\right),\\
    ds^{2}_{\mathrm{AdS}_{5}}& = \frac{1}{z^{2}}\bigg(-dt^{2} +dz^{2}+\sum_{i=1}^{3}(dx^{i})^{2}\bigg)
    \end{aligned}
\end{equation}
where the following commonly used notations have been adopted
\begin{equation}
    \begin{aligned}
       \mu_{1} & = \sin{\beta}\sin{\alpha},\hspace{0.5em}\mu_{2} = \sin{\beta}\cos{\alpha},\hspace{0.5em} \mu_{3} = \cos{\beta} \\
         \sum_{i=1}^{3}\mu^{2}_{i} & = 1, \quad \sum_{i=1}^{3} d\phi_{i} = d\psi      \\
    G^{-1} & = 1 + \hat{\gamma}^{2}\left(\mu^{2}_{1}\mu^{2}_{2} + \mu^{2}_{2}\mu^{2}_{3} + \mu^{2}_{1}\mu^{2}_{3}\right),\hspace{1em} \hat{\gamma} = R^{2} \gamma,
    \end{aligned}
\end{equation}
and in what follows we assume $R=1$.

For the Type IIB superstring on this background in the Metsaev--Tseytlin formalism (where the above field configuration is the part of the supercoset $\frac{SU(2,2|4)}{SO(1,4)\times SO(5)}$) the explicit form of the Lax connection is known \cite{Frolov:2005dj}. Since this is not the case for other examples, we would like to demonstrate that the deformation does not break closed curves in the Poincaré sections, indicating regular (non-chaotic) dynamics of the system. At least for a particular embedding, which is however chosen with no fine-tuning behind. Hence, we consider the following embedding ansatz
\begin{equation}\label{eq:KAMAnsatz}
    \begin{aligned}
         \phi_{1} &= \phi_{1}(\sigma) = \lambda_{1}\,\sigma, \\
        \phi_{2} &= \phi_{2}(\sigma) = \lambda_{2}\,\sigma, && & \beta & = \beta(\tau), \\
        \phi_{3}& = \phi_{3}(\sigma) = \lambda_{3}\,\sigma, && & \alpha & = \alpha(\tau) ,  
    \end{aligned}
\end{equation}
where $\lambda_{1},\lambda_{2},\lambda_{3}\in \mathbb{Z}$ have the meaning of winding numbers of the string along the $\mathrm{U}(1)\times \mathrm{U}(1) \times \mathrm{U}(1)$ isometric direction of the $\SS^5$, and $\t,\s$ denote world-sheet coordinates. The idea of such a class of ans\"atze, commonly used in the literature,  is to turn off oscillatory modes of the string thus effectively reducing its dynamics to a point-like object, while keeping information of the non-trivial cycles of the background manifold by keeping track of winding numbers.

The light cone coordinates are chosen to be completely in the AdS space: $x^{\pm} = \frac{1}{\sqrt{2}}\big(t \pm x^{3}\big)$, and the light cone gauge is fixed by $x^{-} = \tau$. We assume that in the AdS space the string always sits at the point $z=1$, $x^1=0$, $x^2=0$, that gives the condition $p_{i} = 0$, $i=1,2,z$. Indeed, from the form of the full Hamiltonian 
\begin{equation}
    \begin{aligned}
    \mc{H} & = z^2 \left(p_1{}^2+p_2{}^2+p_3{}^2\right)+ p_{\alpha }{}^2 \csc^2\beta  +p_{\beta }{}^2+ \left(\lambda _1^2 \mu _1^2+\lambda _2^2 \mu _2^2+\lambda _3^2 \mu _3^2\right)
    \end{aligned}
\end{equation}
one concludes that for $p_{i} = 0$ the direction $z$ is flat. Note that the the deformation parameter $\g$ drops from the Hamiltonian, that however does not mean that dynamics of the string does not feel the deformation. Indeed, in this RNS picture we do not take into account the R-R sector that also changes with the deformation. To keep track of these changes one has to work in the Green--Schwarz formalism, where, for examples, the corresponding Lax connection changes. In our further examples for tri-vector deformations dependence on deformation parameter can be seen already in the NS-NS sector, and hence this approach is sufficient for our goals. Therefore, let us proceed to demonstrate plots of Poincar\'e sections for this integrable model.

Substituting explicit expressions for $\mu_i$ we find that  the final expression reads
\begin{equation}
    \begin{aligned}
        \mathcal{H} = p_{\beta}^{2} + \frac{p_{\alpha}^{2}}{\sin^{2}\beta} + \lambda_{1}^{2}\sin^{2}\beta\sin^{2}\alpha + \lambda_{2}^{2}\sin^{2}\beta\cos^{2}\alpha + \lambda_{3}^{2}\cos^{2}\beta.
    \end{aligned}
\end{equation}
 Hamiltonian equations of motion read
\begin{equation}
    \begin{aligned}
        \dot{\alpha} &= \frac{2p_{\alpha}}{\sin^{2}\beta}, && & \dot{p_{\alpha}} &= \left(\lambda_{2}^{2} -\lambda_{1}^{2}\right)\sin{2\alpha}\sin^{2}\beta
        \\
        \dot{\beta} &= 2p_{\beta} && & 
        \dot{p_{\beta}} &= \left(\lambda_{3}^{2} -\lambda_{1}^{2}\sin^{2}\alpha - \lambda_{2}^{2}\cos^{2}\alpha\right)\sin2\beta + \frac{2p_{\alpha}^{2}}{\sin^{2}\beta}\cot\beta.
    \end{aligned}    
\end{equation}
Level matching condition requires $p_{\phi_1}=p_{\phi_2}=p_{\phi_3} = 0$, that also follows from equations of motion. 

\begin{figure}[H]
\centering 
    \includegraphics[width = 0.45\textwidth]{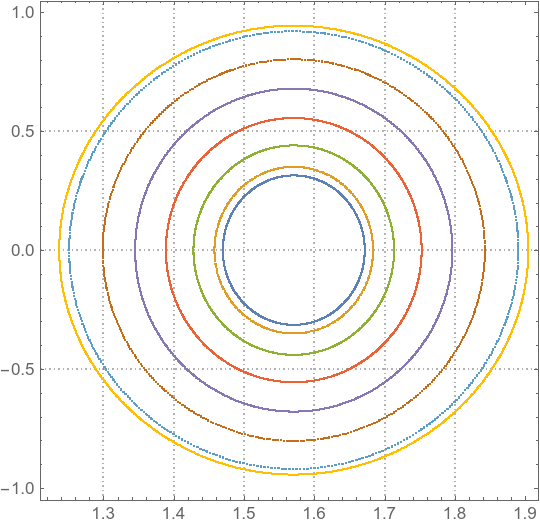}
    \includegraphics[width = 0.45\textwidth]{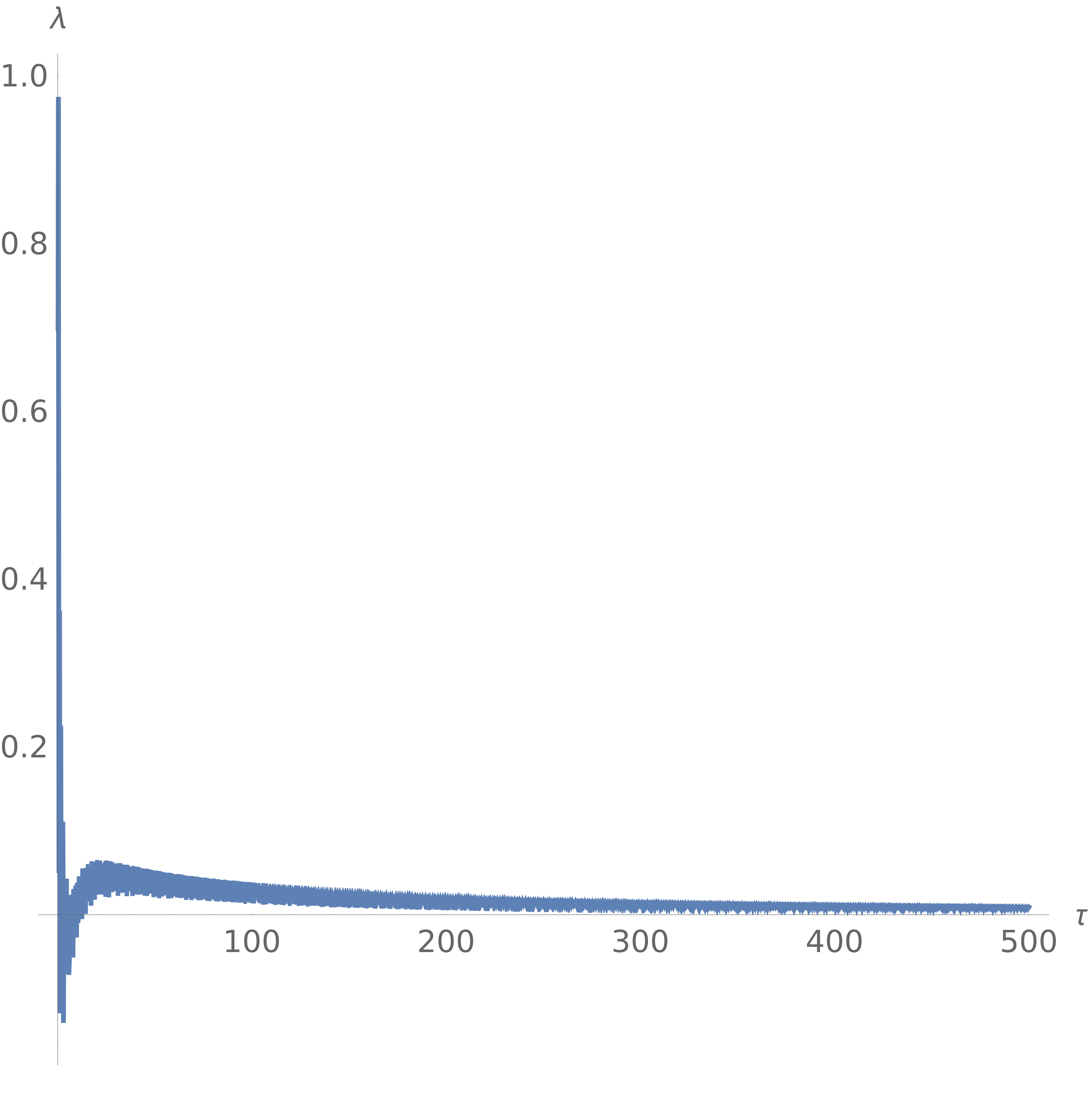}
    \caption{Numerical plots of the Poincare section and Lyapunov exponents for the Lunin-Maldacena of $\mathrm{AdS}_{5}\times\mathbb{S}^{5}$. The energy of the string $E = 3\p/4$. }
    \label{eq:ads5xs5}
\end{figure}

Now we need to solve the resulting system of equations. To do this, we need to set the initial data and also take into account that the energy $E$ is an integral of motion. As a result, we obtain the following dependence between the initial conditions
\begin{equation}
    p_{\alpha}(0) = \sin\beta(0) \sqrt{E^{2} - p_{\beta}^{2}(0) - \lambda_{1}^{2}\sin^{2}\beta(0)\sin^{2}\alpha(0) - \lambda_{2}^{2}\sin^{2}\beta(0)\cos^{2}\alpha(0) - \lambda_{3}^{2}\cos^{2}\beta(0)}    
\end{equation}
For our analysis, we set $\lambda_{1} = 1$, $\lambda_{2} = 2$, $\lambda_{3} = 3$, and use the following initial conditions:
\begin{equation*}
    \begin{gathered}
        \alpha(0) = \frac{3\pi}{2}, \quad \beta(0)\in\left[\frac{\pi}{2}, \frac{\pi}{2} + \frac{\p}{10}\right], \quad p_{\beta}(0) = \fr{\p}{10}.
    \end{gathered}
\end{equation*}
Energy of the string is taken to be $E=3\p/4$. In order to calculate the Lyapunov exponent, we choose to work with the initial conditions $E = 3\pi/4$ together with  \{$\alpha(0) = \frac{3\pi}{2}$, $\beta(0) = \frac{\pi}{2} + \frac{\pi}{20}$, $p_{\beta}(0) = \frac{\pi}{10}$\}. Here we take initial condition in between of those limits chosen to draw Poincar\'e sections since the plot for Lyapunov exponent does not change its form within the range.

From the presented plots we conclude, that the Poincaré section and the Lyapunov exponent show the behaviour typical for an integrable system. Decay of the Lyapunov exponent with time shows that trajectories do not diverge with time and the form of the Poincar\' sections indicates that dynamics of the system is regular, at least in the chosen sector. This is certainly consistent with the analytical result showing classical integrability of this system.

\section{Tri-vector deformed backgrounds}
\label{sec:trivec}

\subsection{Abelian internal deformation of \texorpdfstring{$\mathrm{AdS}_{4}\times \CP^3$}{AdS4xCP3}}

We now turn to the main goal that is analysis of invariant tori and Lyapunov exponents for the Type II(A) sigma model on tri-vector deformed backgrounds. As before we will be dealing only with bosonic sector of the model under a certain embedding ansatz.  As our first example we take an abelian $\rmU(1)^3$ deformation of the $\AdS_4 \times \CP^3$ background of the Type IIA string. Integrability of the superstring on the undeformed $\AdS_4 \times \CP^3$ has been shown explicitly by constructing a zero-curvature Lax pair: in \cite{Arutyunov:2008if,Stefanski:2008ik} for the reduced description by the super-coset $\fr{\mathrm{OSp}(6|4)}{\rmU(3) \times \rmSO(1, 3)}$ and in \cite{Sorokin:2010wn} for sectors of the model outside of the supercoset. 

Since we are working with a specific embedding of the superstring our analysis is coarse enough to not to feel the difficulties related to the super-coset description and the amount of supersymmetries of the model. Not to mention that the deformation in question preserves only $\mc{N}=1$ SUSY. The idea is to start with  the Lunin--Maldacena deformation \cite{Lunin:2005jy} of the $\AdS_4\times \SS^7$ solution to 11D supergravity equations and to reduce it along the undeformed Hopf cycle to get the $\CP^3$ from the 7-sphere. Hence, we start with the following 11-dimensional field configuration 
\begin{equation}
    \begin{aligned}
        ds^{2}_{11}  & = G^{-1/3}R^{2}\left(\frac{1}{4}ds^{2}_{\mathrm{AdS}_{4}} + \sum_{i=1}^{4}\left(d\mu^{2}_{i} + G\mu^{2}_{i}d\phi^{2}_{i}\right) + 16\hat{\gamma}^{2}G\mu^{2}_{1}\mu^{2}_{2}\mu^{2}_{3}\mu^{2}_{4}\left(\sum_{i=1}^{4} d\phi_{i}\right)^{2}\right),
        \\
        ds^{2}_{\mathrm{AdS}_{4}}& = \frac{1}{z^{2}}\bigg(-dt^{2} +dz^{2}+\sum_{i=1}^{2}(dx^{i})^{2}\bigg), 
        \\
        \mathcal{F}_4  & = \frac{3}{8}\bigg(\omega_{AdS} + 16\hat{\gamma}\sin^{5}\theta\cos\theta\sin^{2}2\alpha\sin2\beta\hspace{0.2em}d\theta\wedge d\alpha\wedge d\beta\wedge d\psi\bigg) +
        \\
        & + \hat{\gamma}R^{3}d\Bigg[G\bigg(\prod_{i=1}^{4}\mu_{i}^{2}\bigg)\bigg(\frac{d\phi_{123}}{\mu_{4}^{2}} + \frac{d\phi_{134}}{\mu_{2}^{2}} - \frac{d\phi_{124}}{\mu_{3}^{2}} - \frac{d\phi_{234}}{\mu_{1}^{2}}\bigg) \Bigg],
    \end{aligned}
\end{equation}
where $d\phi_{abc} = d\f_a\wedge d\f_b \wedge d \f_c$ and as before the following notations have been introduced
\begin{equation}
    \begin{gathered}
    \mu_{1} = \sin{\beta}\sin{\alpha}\sin{\theta},\hspace{0.5em} \mu_{2} = \sin{\beta}\sin{\alpha}\cos{\theta},\hspace{0.5em} \mu_{3} = \sin{\beta}\cos{\alpha},\hspace{0.5em} \mu_{4} = \cos{\beta}
        \\
        \Delta = \mu^{2}_{1}\mu^{2}_{2}\mu^{2}_{3}\mu^{2}_{4}\sum_{i=1}^{4}\mu_{i}^{-2},\hspace{1em} G^{-1} = 1 + \hat{\gamma}^{2}\Delta, \hspace{1em} \hat{\gamma} = \gamma R^{3}   
        \\
        \sum_{i = 1}^{4} \mu_{i}^{2} = 1,\hspace{1em} \sum_{i = 1}^{4}d\phi_{i} = d\psi.
    \end{gathered}
\end{equation}
The reduction from eleven to ten dimensions is performed according to the standard Kaluza--Klein ansatz, that for the metric reads 
\begin{equation}
    g^{(11)}_{MN} = 
    \begin{bmatrix}
        e^{-\fr23 \Phi}g^{(10)}_{mn} + e^{\fr43 \Phi}C_{m}C_{n} & e^{\fr43 \Phi}C_{m}
        \\
       e^{\fr43 \Phi}C_{n} & e^{\fr43 \Phi}
    \end{bmatrix}.
\end{equation}
Here $\Phi$ is the dilaton and $C_m$ is the R-R 1-form field of the Type IIA theory. 

To derive the Kalb--Ramond $B$-field we start with the following 3-form gauge field in 11 dimensions  \cite{Bakhmatov:2019dow} 
\begin{equation}
    \begin{aligned}
         C =&\  C_0 + \gamma R^{6}G\Big(s_\q^{4}c_\q^{2}s_\a^{2}c_\s^{2}c_\b^{2} \,d\phi_{1}\wedge d\phi_{2}\wedge d\phi_{3} -  s_\q^{4}c_\q^{2} s_\a^{2} c_\a^{2} s_\b^{2} \,d\phi_{1}\wedge d\phi_{2}\wedge d\phi_{4} +\\
         &+ s_\q^{4} s_\a^{4} c_\q^{2} s_\b^{2} c_\b^{2} \,d\phi_{1}\wedge d\phi_{3}\wedge d\phi_{4} - s_\q^{6} s_\a^{4} c_\a^{2} s_\b^{2} c_\b^{2} \,d\phi_{2}\wedge d\phi_{3}\wedge d\phi_{4} \Big)  ,
    \end{aligned}
\end{equation}
where $dC_0 = F_4$ gives the undeformed 4-form flux. The latter has legs only along $\AdS_4$ and hence when reduced the $C_0$ does not contribute to the Kalb--Ramond field. Since the NSR formalism for the classical bosonic string does not involve RR fields, we will need only the B-field, for which we have
\begin{equation}
    \begin{aligned}
        B & = \gamma  G \mu_1^2 \mu_2^2 d\phi_3\wedge d\theta_1-\gamma  G \mu_1^2 \mu_3^2 d\phi_3\wedge d\theta_2+\gamma  G \mu_3^2 \mu_2^2  d\theta_1\wedge d\theta_2.
    \end{aligned}
\end{equation}

As in the previous section, we proceed to the light cone coordinates $x^{\pm} = \frac{1}{\sqrt{2}}\left(t \pm x^{1}\right)$ assuming that the string is at the point $x_{2} = 0$, $\alpha = \frac{\pi}{2}$ and choose the same ansatz \eqref{eq:KAMAnsatz} for windings. As before  level matching requires $p_{\phi_1}=p_{\phi_2}=p_{\phi_3} = 0$, the directions $x_2$ and $z$ are flat given $p_{2} = 0, p_{z} = 0$ and the same is for $\a$ if $p_{\alpha} =0$. The resulting Hamiltonian takes the following form
\begin{equation}
    \begin{aligned}
        \mathcal{H} & = p_{\theta}^{2} + \frac{p_{\beta}^{2}}{\sin^{2}{\theta}} + \frac{1}{16}\Bigg(\lambda_{1}^{2}\bigg(5 - 3\cos{4\beta} - 2\cos^{2}{2\beta}\cos{4\theta} - 8\cos{2\theta}\sin^{2}{2\beta}\bigg)  \\
        & + \frac{\lambda_{2}^{2}}{4}\bigg(7 + 4\cos{2\beta} - 3\cos{4\beta} - 8\cos^{4}{\beta}\cos{4\theta} - 8\cos{2\theta}\sin^{2}{2\beta}\bigg) + 4\lambda_{3}^{2}\sin^{2}{2\theta} 
        \\
        & - 8\lambda_{2}\lambda_{3}\cos^{2}{\beta}\sin^{2}{2\theta} + 32\lambda_{1}\lambda_{2}\bigg(\cos{2\beta}\sin^{2}{\theta} - 1\bigg)\cos^{2}{\beta}\sin^{2}{\theta} + 8\lambda_{1}\lambda_{3}\cos{2\beta}\sin^{2}{2\theta}\Bigg).
    \end{aligned}
\end{equation}
The corresponding equations of motion read 
\begin{equation}
\label{eq:ppp_cp3}
    \begin{aligned}
         \dot{\theta} & = 2p_{\theta} 
         \\
         \dot{\beta} & =  \frac{2p_{\beta}}{\sin^{2}{\theta}}
         \\
         \dot{p}_{\theta}  &= \frac{2p_{\beta}^{2}}{\sin^{2}{\theta}}\cot{\theta} - \frac{\left(\lambda_{2} - 2\lambda_{1}\right)^{2}}{4}\sin^{2}{2\beta}\sin{2\theta} -\frac{\sin{4\theta}}{2}\bigg(\lambda_{3} - \lambda_{2}\cos^{2}{\beta} + \lambda_{1}\cos{2\beta}\bigg)^{2}
         \\
         \dot{p}_{\beta} & = \left(\lambda_{2} - 2\lambda_{1}\right)\sin{2\beta}\sin^{2}{\theta}\bigg(2\lambda_{1}\cos{2\beta}\sin^{2}{\theta} + \lambda_{2}\cos{2\theta}\cos^{2}{\beta} + \lambda_{2}\sin^{2}{\beta} - 2\lambda_{3}\cos^{2}{\theta}\bigg)
    \end{aligned}
\end{equation}

\subsubsection*{Numerical results}

\begin{figure}[H]
    \centering
        \begin{subfigure}{0.4\textwidth}
            \centering
             \includegraphics[width = 0.9\textwidth]{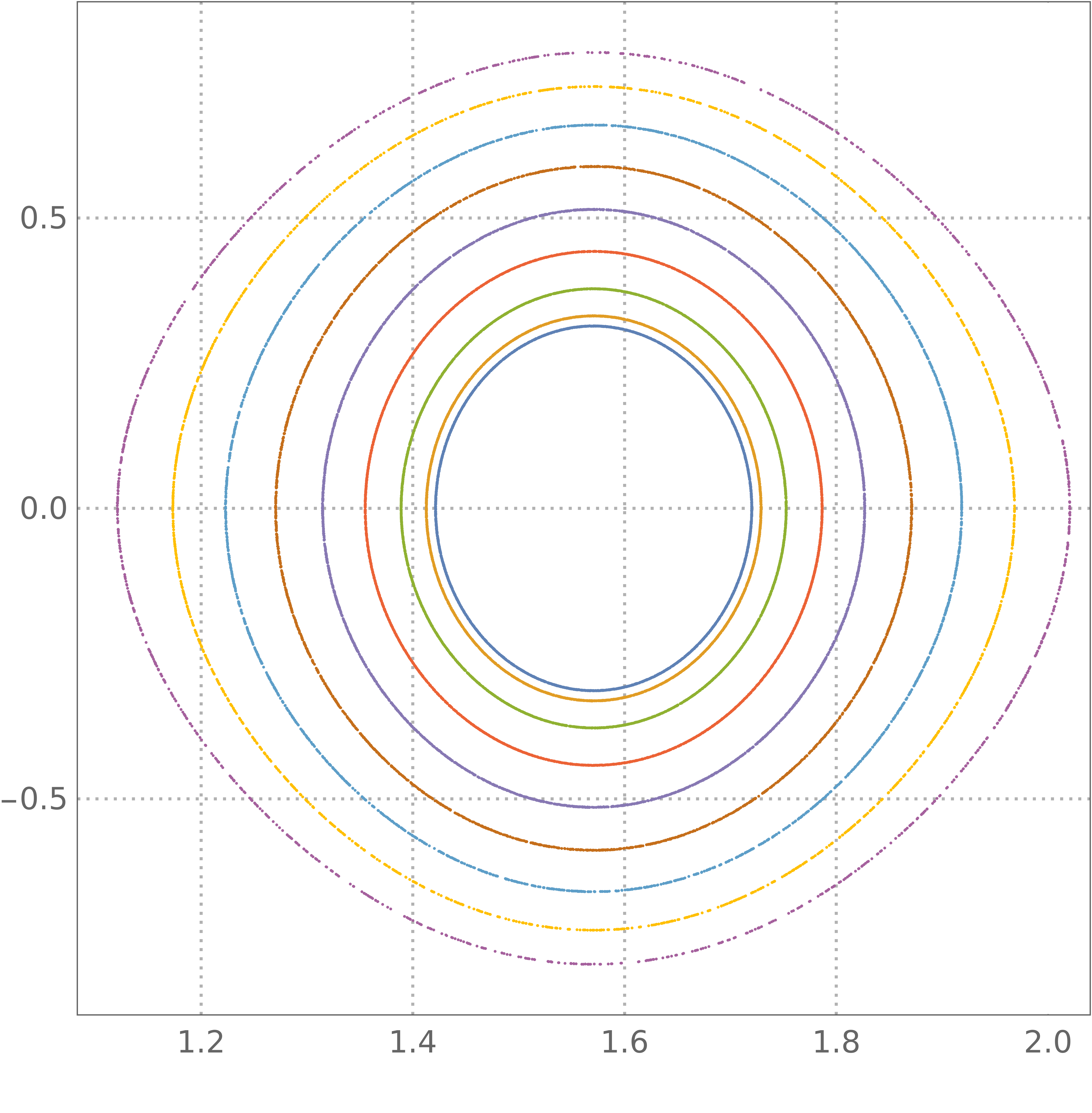}
             \caption{$(\theta,p_{\theta})$}
        \end{subfigure}    
        \begin{subfigure}{0.4\textwidth}
            \centering
             \includegraphics[width = 0.9\textwidth]{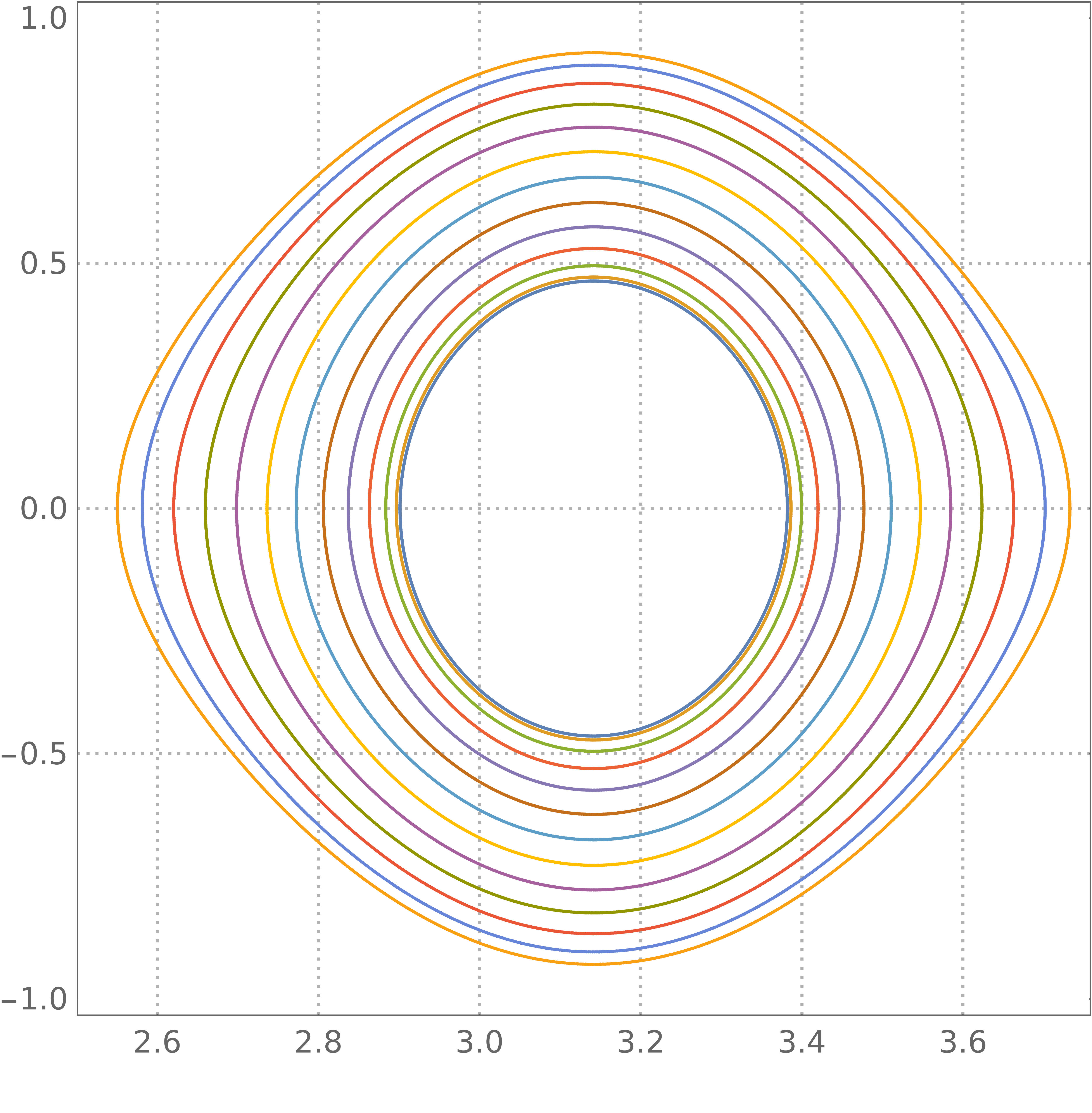}
             \caption{$(\beta,p_{\beta})$}
        \end{subfigure}   
        \\
        \begin{subfigure}{0.4\textwidth}
            \centering
               \includegraphics[width = 0.9\textwidth]{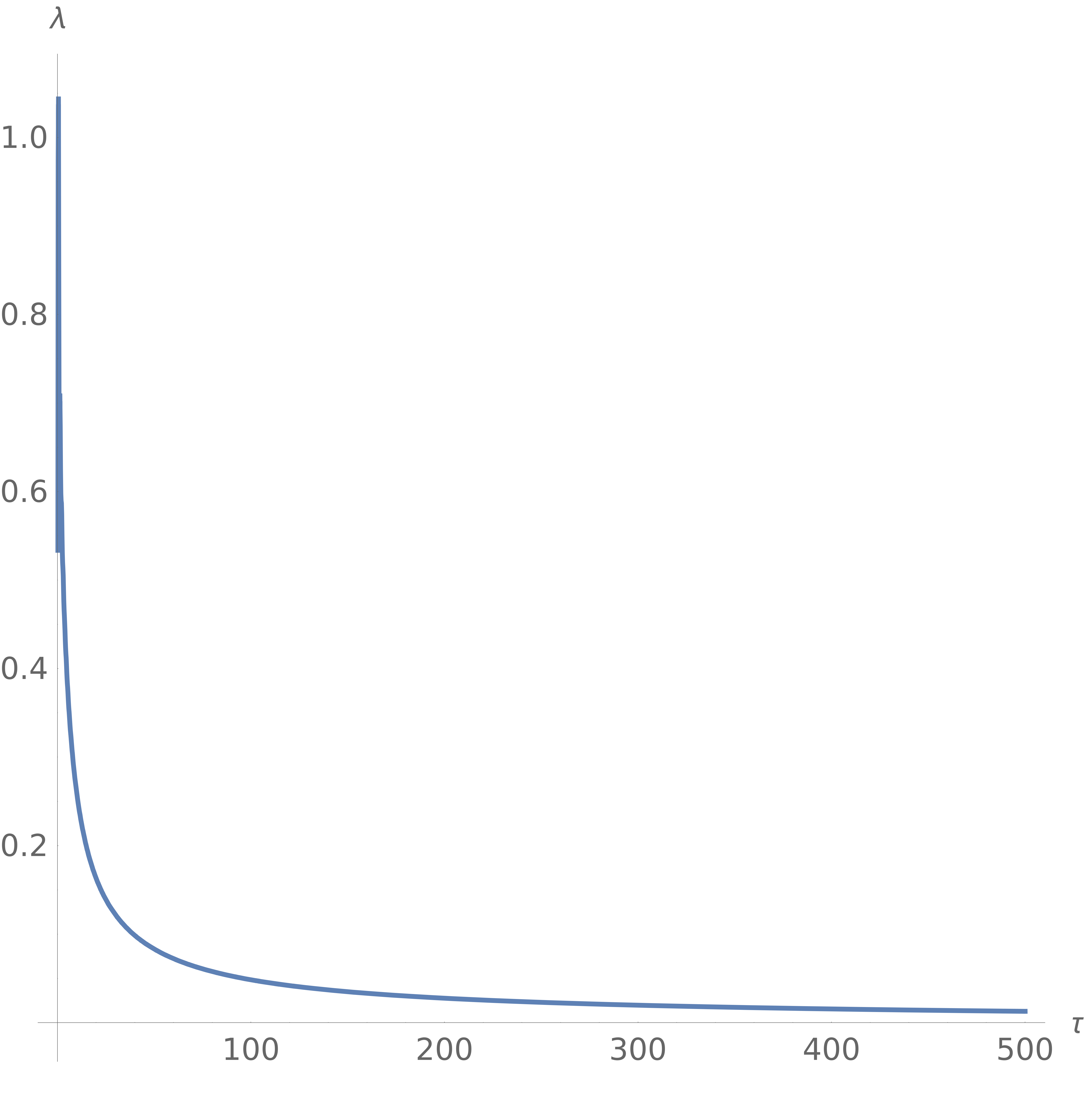}
        \end{subfigure}
        \begin{subfigure}{0.4\textwidth}
            \centering
               \includegraphics[width = 0.9\textwidth]{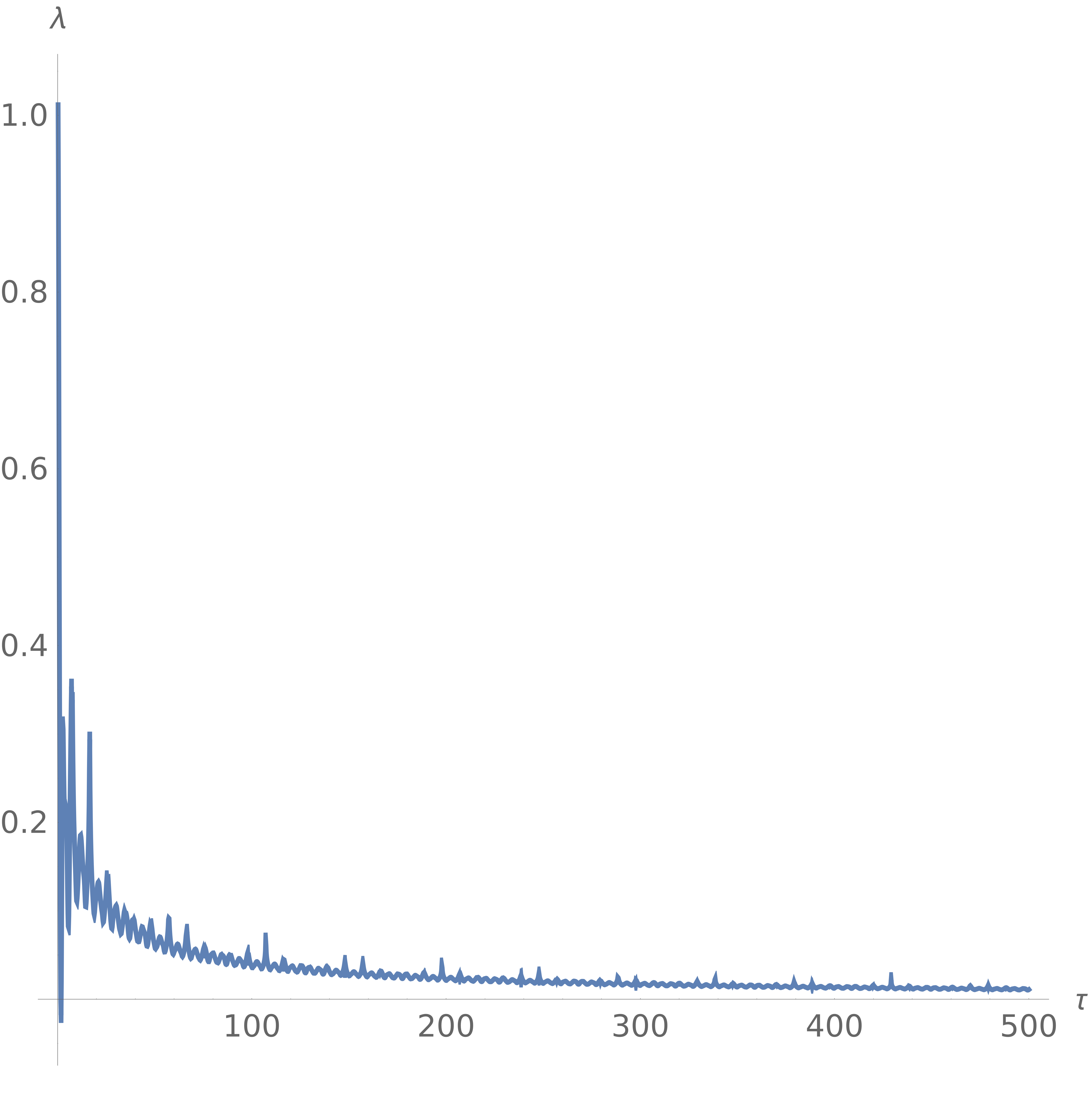}
        \end{subfigure}
    \caption{Numerical plots of the Poincar\'e section and  Lyapunov exponents for abelian internal deformation of $\mathrm{AdS}_{4}\times\mathbb{C}\mathrm{P}^{3}$}
    \label{fig:ads4xcp3_abelian}
\end{figure}

Let us now proceed with numerical analysis of the obtained system. Conservation of energy imposes the following restriction of the initial conditions
\begin{equation}
    \begin{gathered}
        p_{\beta}(0) = \frac{\sin{\theta(0)}}{8}\Bigg(
        64\left(E - p_{\theta}(0)^{2}\right)  + 32 \lambda_{2}\lambda_{3} \cos^{2}{\beta(0)}\sin^{2}{2\theta(0)}  + 8\lambda_{3}^{2}\Big(\cos{4\theta(0)} - 1\Big) +
        \\
        + \lambda_{2}^{2} \Big(-7 - 4\cos{2\beta(0)} + 3\cos{4\beta(0)} + 8\cos^{4}{\beta(0)}\cos{4\theta(0)} + 8\cos{2\theta(0)} \sin^{2}{2\beta(0)}\Big) + 
        \\
        + 4\lambda_{1}\Big(8\lambda_{2}\cos^{2}{\beta(0)}\sin^{2}{\theta(0)}\left(4 - 4 \cos{2\beta(0)}\sin^{2}{\theta(0)}\right) - 8\lambda_3\cos{2\beta(0)}\sin^{2}{2\theta(0)}\Big) + \\
        + 4\lambda_{1}^{2}\Big(-5 + 3\cos{4\beta(0)} + 2\cos^{2}{2\beta(0)}\cos{4\theta(0)} + 8\cos{2\theta(0)}\sin^{2}{2\beta(0)}\Big)\Bigg)^{1/2}.
    \end{gathered}
\end{equation}
Poincar\'e sections for the model are presented in Fig. \ref{fig:ads4xcp3_abelian}. These plots correspond to two set of embeddings and two choices of position of the Poincar\'e plane: $(\theta,p_{\theta})$ and$(\beta,p_{\beta})$. The initial data and windings are chosen as follows:
\begin{enumerate}
    \item $(\theta,p_{\theta})$
    \begin{equation}
        \begin{gathered}
            \lambda_{1} = 1,\hspace{1em}\lambda_{2} = 2 ,\hspace{1em}\lambda_{3} = 3
            \\
            \beta(0) = \frac{\pi}{2},\hspace{1em}\theta(0) \in \left[\frac{\pi}{2},\frac{\pi}{2}+\frac{\pi}{8}\right],\hspace{1em}p_{\theta}(0) = \frac{\pi}{10},\hspace{1em} E = \frac{\pi}{5}
        \end{gathered}
    \end{equation}
    \item $(\beta,p_{\beta})$
    \begin{equation}
        \begin{gathered}
            \lambda_{1} = \lambda_{2} = \lambda_{3} = 1
            \\
            \beta(0) \in\left[\pi, \pi + \frac{\pi}{6}\right],\hspace{1em}\theta(0) = \frac{\pi}{2},\hspace{1em}p_{\theta}(0) = \frac{\pi}{10},\hspace{1em} E= \frac{\pi}{10}
        \end{gathered}
    \end{equation}
\end{enumerate}
From \eqref{eq:ppp_cp3} we see that equations for the momenta significantly simplify if the winding numbers are chose as $\lambda_{1} = 1$, $\lambda_{2} = 2$, $\lambda_{3} = 3$, hence this is our choice for the first plot. The second plot is to show a different section with different winding numbers.

As before we observe, that intersection points belong to  closed curves meaning that phase curves lie on tori. Since the deformation parameter drops from the Hamiltonian in the NS-NS sector the tori are (trivially) invariant under the deformation. This will not be the case for non-abelian deformations. Lyapunov exponent $\l$ showing divergence of two phase curves is presented on the bottom plot of Fig. \ref{fig:ads4xcp3_abelian} and has been drawn for the following initial conditions
\begin{enumerate}
    \item
    \begin{equation}
        \begin{gathered}
            \lambda_{1} = 1,\hspace{1em}\lambda_{2} = 2 ,\hspace{1em}\lambda_{3} = 3
            \\
            \beta(0) = \frac{\pi}{2},\hspace{1em}\theta(0) = \frac{9\pi}{16},\hspace{1em}p_{\theta}(0) = \frac{\pi}{10},\hspace{1em} E = \frac{\pi}{5}
        \end{gathered}
    \end{equation}
    \item 
    \begin{equation}
        \begin{gathered}
            \lambda_{1} = \lambda_{2} = \lambda_{3} = 1
            \\
            \beta(0) = \frac{7\pi}{6},\hspace{1em}\theta(0) = \frac{\pi}{2},\hspace{1em}p_{\theta}(0) = \frac{\pi}{10},\hspace{1em} E= \frac{\pi}{10}
        \end{gathered}
    \end{equation}
\end{enumerate}
We again observe the characteristic behaviour of an integrable system.

To our knowledge, no analytical results have been presented in the literature for such a tri-vector deformed sigma-model, however it must be possible to generalize the approach of \cite{Orlando:2019rjg} where preservation of Lax connection under a general global $\rmO(d,d)$ transformation was shown. An argument would be the following. On one hand, following \cite{Sakamoto:2017cpu} one concludes that deformation with constant parameter are global coordinate transformations in the extended space of the corresponding exceptional field theory. On the other hand, given a formulation of the string dynamics covariant w.r.t. global exceptional duality group, similar to \cite{Arvanitakis:2018hfn}, one concludes that equations of motion are also covariant and hence their representation in terms of Lax connection will also be covariant.

\subsection{Non-abelian external deformation  of \texorpdfstring{$\mathrm{AdS}_{4}\times\mathbb{C}\mathrm{P}^{3}$}{AdS4xCP3}}

Certainly, preservation of regular tori under an abelian tri-vector deformation of an integrable sigma-model is a pretty degenerate example since the generalized Yang--Baxter equation holds trivially due to vanishing structure constants. More relevant information can be gained from analysis of non-abelian deformations, that we proceed with immediately. Therefore, our second example is a non-abelian deformation of $\mathrm{AdS}_{4}\times\mathbb{C}\mathrm{P}^{3}$ along the AdS directions similar to that of \cite{Bakhmatov:2020kul}. We follow the same algorithm as before and start with the PPM deformation of the 11-dimensional background $\mathrm{AdS}_{4}\times\mathbb{S}^{7}$ found in \cite{Bakhmatov:2020kul} and reduce it to ten dimensions. The corresponding tri-vector  has the following form
\begin{equation}
    \begin{aligned}
        \W & = \frac{4 }{R^3}\rho_a x^a \dt_0\wedge \dt_1 \wedge \dt_2,
    \end{aligned}
\end{equation}
where $\rho_{a}$ denote deformation parameters with $a = 0,1,2$. This deformation has been dubbed PPM since the generators are taken to be along momenta and boosts. The resulting Type IIA background reads
\begin{equation}
\label{eq:ads4xcp3def}
    \begin{aligned}
        ds^2 & = \fr{R^3}{4 \sqrt{z^3 - \r_a x^a}\sqrt{z}}\Big(-(dx^0)^2 +(dx^1)^2 +(dx^2)^2+\fr{z^3 - \r_a x^a}{z^3}dz^2 \Big) +\fr{R^3\sqrt{z^3 - \r_a x^a}}{z^{\fr32}}ds_{\CP^3}^2, \\
        e^{2\Phi} & = \fr{R^3\sqrt{z^3 - \r_a x^a}}{z^{\fr32}},
    \end{aligned}
\end{equation}
where the interval $ds_{\CP^3}^2$ on $\CP^3$ is written in the parametrization \eqref{eq:cp3two}. The R-R fields are given by the following expressions  
\begin{equation}
\label{eq:ads4xcp3def_RR}
    \begin{aligned}
        C_1 & = \frac{1}{2} c_{\theta _1} c^2_\xi d\phi_2+\frac{1}{2} c_{\theta_2} s^2_{\xi}d\phi_3+\frac{1}{2} c_{2 \xi}d\phi_1 ,\\
        C_3 & = -\fr{R^3}{8 (z^3 - \r_a x^a)} dx^0\wedge dx^1 \wedge dx^2.
    \end{aligned}
\end{equation}
This background is a solution to supergravity equations for any choice of $\r_a$, i.e. generalized classical Yang--Baxter equation has been manifestly taken into account. 

Since the deformation parameters $\r_1$ and $\r_2$ enter in a symmetric way as $\rho_a x^a$, it does not really matter which of the two is non-zero and for simplicity we choose $\r_1 =0$. The case of $\r_0$ is more tricky as it introduces time dependence into the Hamiltonian. We postpone discussion of this case to a further section, and here choose  $\rho_{0} = 0$. Let us also denote $\rho_{2} = \gamma$, where $\g$ will be varied. We turn to the light-cone coordinates as before and it appears consistent to set $\q_2=0$, $p_{\q_2}=0$. From the form of the resulting Hamiltonian 
\begin{equation}
    \begin{gathered}
        \mathcal{H} = \frac{1}{4 z^{3/2} \sqrt{z^3-\gamma  x_2}} \bigg[-16 \gamma  p_2^2 x_2 z^2+16 p_2^2 z^5+16 z^5 p_z^2+16 z^3 \sec ^2(\xi ) p_{\theta _1}^2+4 z^3 p_{\xi }^2 + 
        \\ 
        + \gamma  \lambda _3^2 x_2 \sin ^4(\xi )-\gamma  \lambda _3^2 x_2 \sin ^2(\xi )-\lambda _2^2 \cos ^2(\xi ) \left(\cos ^2\left(\theta _1\right) \cos ^2(\xi )-1\right) \left(z^3-\gamma  x_2\right) +
        \\
        + \lambda _1 \sin ^2(2 \xi ) \left(\lambda _2 \cos \left(\theta _1\right)-\lambda _3\right) \left(z^3-\gamma  x_2\right)-\frac{1}{2} \lambda _2 \lambda _3 \cos \left(\theta _1\right) \sin ^2(2 \xi ) \left(z^3-\gamma  x_2\right) +
        \\
        +\lambda _1^2 \sin ^2(2 \xi ) \left(z^3-\gamma  x_2\right)-\lambda _3^2 z^3 \sin ^4(\xi )+\lambda _3^2 z^3 \sin ^2(\xi )\bigg],
    \end{gathered}
\end{equation}
we conclude that neither $z$ and $x_2$ direction are flat nor it is consistent to set $z$ and $x_2$ to be constant. Therefore, one is not able to analyse the system as before by focusing at angle variables and their corresponding momenta. 

To gain more understanding about the system let us first restrict ourselves to a particular class of embeddings, where Poincar\'e sections for the variables $\x$ and $p_\x$ can be obtained analytically. For that we set $\l_1=1$, $\l_2=0$, $\l_3=2$ and the Hamiltonian simplifies drastically
\begin{equation}
    \mc{H}=\frac{\sqrt{z}}{\sqrt{z^3-\gamma  x_2}} \bigg(-4 \gamma  p_2^2 x_2+4 \left(p_2^2+p_3^2\right) z^3+z \left(4 p_{\theta _1}^2 \cos^{-2}\xi \,  +p_{\xi }^2\right)\bigg).
\end{equation}
One notices immediately that equations of motion give $\dot{p}_{\q_1} = 0$ rendering $p_{\q_1} = $ const. Next, from equations of motion for the pair $(\x,p_\x)$, that read
\begin{equation}
    \begin{aligned}
        \dot{\x} & = \fr{2z^{\fr32}}{\sqrt{z^3 - \g x_2}} p_\x, &&
        \dot{p}_\x & = -\fr{8z^{\fr32}}{\sqrt{z^3 - \g x_2}} \fr{\sin \x}{\cos^3 \x}p_{\q_1}^{2},
    \end{aligned}
\end{equation}
we find a second integral of motion
\begin{equation}
    p_\x^2 + 4\fr{p_{\q_1}^2}{\cos^2 \x} = 4\a = \const.
\end{equation}
We conclude that the variables $p_\x$ and $\cos^{-1}\x$ on the equations of motion form an ellipse for any value of the other variables. This can be though of as a Poincar\'e section of the system in the plane $(\x,p_\x)$ at any given point.  

The equation for $\q_1$, that is 
\begin{equation}
    \dot{\q}_1 = \fr{8z^{\fr32}}{\sqrt{z^3 - \g x_2}} p_{\q_1}\frac{1}{\cos^2{\xi}},
\end{equation}
can be completely integrated if combined with $\dot{\x}$ taking into account the expression for $p_\x$. This gives
\begin{equation}
    \tan(\q_1 + \bar{\q}_1) = \fr{2 \sin \x}{p_\x},
\end{equation}
where $\bar{\q}_1$ is an integration constant. Therefore, all angle variables behave  regularly as functions of each other and Poincar\'e sections in the $(\x,p_\x)$ plane are as in the previous cases. 

\begin{figure}[H]
    \centering
        \begin{subfigure}{0.4\textwidth}
            \centering
             \includegraphics[width = 0.9\textwidth]{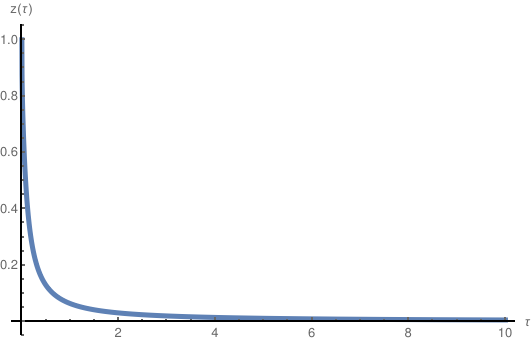}
             \caption{$z=z(\t)$}
        \end{subfigure}    
        \begin{subfigure}{0.4\textwidth}
            \centering
             \includegraphics[width = 0.9\textwidth]{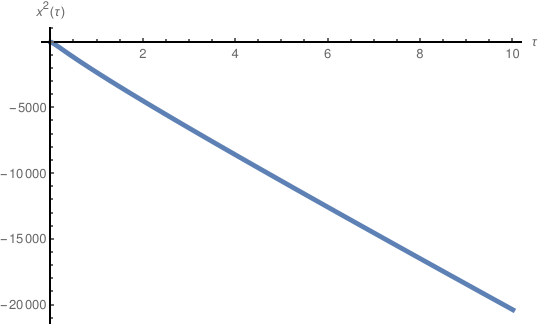}
             \caption{$x^2=x^2(\t)$}
        \end{subfigure}   
    \caption{Numerical plots of the AdS coordinates $x^2$ and $z$ of the string for winding numbers $\l_1=1$, $\l_2=0$, $\l_3=2$.}
    \label{fig:zx}
\end{figure}

Certainly, one is also interested in the full dynamics that includes dependence on the time $\t$ of the angle variables $\x$ and $\q_1$ as well as solutions for the functions $z=z(\t)$ and $x^2 = x^2(\t)$. Hamiltonian equations for the latter two decoupled from that for the angles and read
\begin{equation}
    \begin{aligned}
        \dot{z} & = \frac{8 p_3 z^{7/2}}{\sqrt{z^3-\gamma  x_2}}, && & \dot{p}_z & = \frac{12 \a \gamma  x_2 z+4 p_3^2 z^3 \left(7 \gamma  x_2-4 z^3\right)-4 p_2^2 \left(z^3-\gamma  x_2\right) \left(4 z^3-\gamma  x_2\right)}{2 \sqrt{z} \left(z^3-\gamma  x_2\right){}^{3/2}} ,\\
        \dot{x}^2 & = 8 p_2 \sqrt{z} \sqrt{z^3-\gamma  x_2}, &&  & \dot{p}_2 & = 2\frac{\gamma  \sqrt{z} \left( p_2^2 \left(z^3-\gamma  x_2\right)-z \left(\a+ p_3^2 z^2\right)\right)}{ \left(z^3-\gamma  x_2\right){}^{3/2}}.
    \end{aligned}
\end{equation}
These equations appear to be too complicated to be solved analytically, however, numerical methods deal with them well. It appears that plots of numerical solutions for any values of $\a$ and initial conditions are very similar, therefore as an example for choose $\a=1000$, $\g=100$ and the following initial conditions
\begin{equation}
    \begin{aligned}
        z(0)&=1, && & p_z(0) = -30, \\
        x^2(0)& = 0, && & p_2(0) = 1.
    \end{aligned}
\end{equation}
These conditions mean that initially the string sitting at $(z,x)= (1,0)$ receives a momentum $p_z(0)$ towards $z=0$ (the AdS boundary if $\g=0$) and $p_2(0)$ in the direction of positive $x^2$. This choice has been made since the string with momentum $p_z(0)$ away of the point $z=0$ at some moment of time make a U-turn and moves towards $z$. Direction of the momentum $p_2(0)$ does not matter as long as the combination $z^3 - \g x^2$ under the square root is positive. The numerical solution has been obtained for the parameter $\t\in [0,1000]$, however we plot functions only for $\t < 10$, since $z(t)$ decreases very fast tending to zero. The coordinate $x^2(\t)$ continues its linear descent. The plots are presented on Fig. \ref{fig:zx}

These explicit numerical solutions allow to determine behaviour of the angles with $\t$. Indeed, from the equation for $\dot{\x}$ using expression for $p_\x$ we obtain the following integral equation
\begin{equation}
    \arctan\left[\fr{\sqrt{\a}\sin \x(\t)}{\sqrt{\a \cos^2 \x(\t) -  p_{\q_1}^2}}\right] = \int_0^\t \fr{2\,z(\t')^{\fr32}}{\sqrt{z(\t')^3 - \g \, x_2(\t')}} d\t'.
\end{equation}
Numerical integration of the RHS shows that the integral tends to zero very fast with $\t$, and value of $\x$ is determined by an integration constant. Overall, the $\x$ rapidly decreases with time and stabilizes at some value. Physically this means that energy of the angular motion is transferred to the potential energy of the string that tends to the point $z=0$. We conclude, that the Poincar\'e section in the plane $(p_\x,\x)$ deforms with time, however always being an ellipse.

\begin{figure}[H]
    \centering
        \begin{subfigure}{0.4\textwidth}
            \centering
             \includegraphics[width = 0.9\textwidth]{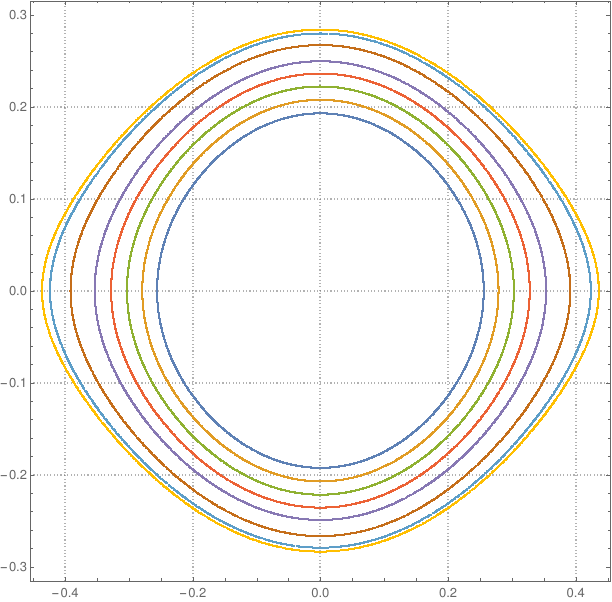}
             \caption{$\g=5$}
        \end{subfigure}    
        \begin{subfigure}{0.4\textwidth}
            \centering
             \includegraphics[width = 0.9\textwidth]{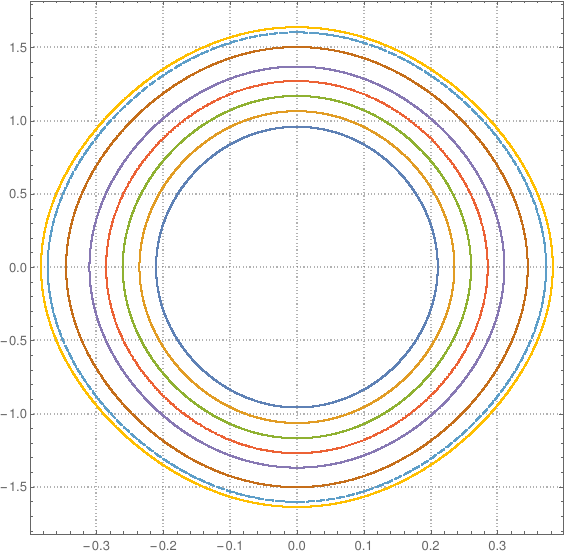}
             \caption{$\g=100$}
        \end{subfigure} \\
        \begin{subfigure}{0.4\textwidth}
            \centering
             \includegraphics[width = 0.9\textwidth]{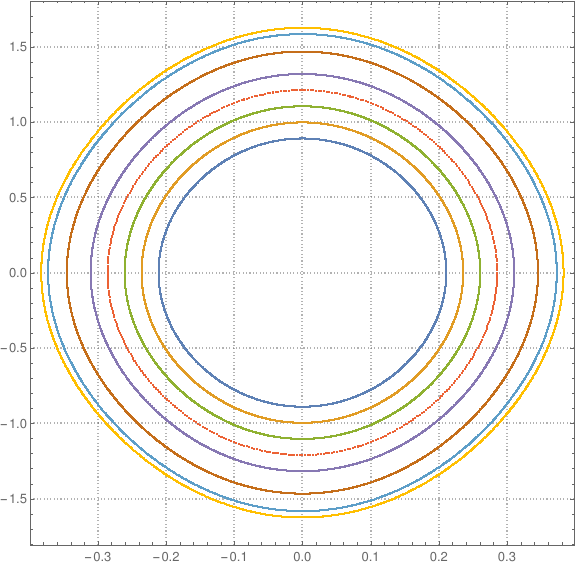}
             \caption{$\g=500$}
        \end{subfigure} 
        \begin{subfigure}{0.4\textwidth}
            \centering
             \includegraphics[width = 0.9\textwidth]{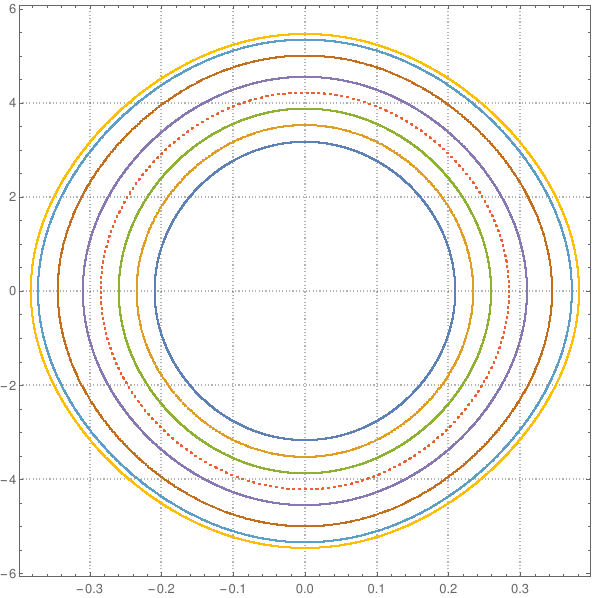}
             \caption{$\g=1000$}
        \end{subfigure} 
    \caption{Poincar\'e sections in the plane $(\x,p_\x)$ for $\l_1=1$, $\l_2=2$, $\l_3=3$ with manually fixed $x^2$ and $z$.}
    \label{fig:ads4xcp3_nonabelian_fixed}
\end{figure}

It seems, that such transfer of energy between $\CP^3$  and the deformed AdS sectors of the strings' is a consequence of ignoring the R-R fields of the background, and on itself is not a definite signature of non-integrability of the model. The first argument for that is explicitly regular dynamics of the angular variables in the example analysed above. As the second example let us suppose that the $z$ and $x^2$ coordinates of the string are somehow fixed (say by a contribution from the R-R sector) to be $z=1$, $x^2=-1$. In this case dynamics of the remaining angular coordinates is regular an Poincar\'e sections are easy to reproduce. For the numerical plots in Fig. \ref{fig:ads4xcp3_nonabelian_fixed} we choose $\l_1=1$, $\l_2=2$, $\l_3=3$ and 
\begin{equation}
    \begin{aligned}
    E&=\frac{\pi }{3}, && \theta_1(0)=0, \\
    \xi(0)&=\left(\frac{\pi }{15},\fr{2 \p}{15}\right), && p_\x(0)=\frac{\pi }{30},
    \end{aligned}
\end{equation}
and $p_{\q_1}$ is determined by the constraints.
The same regular picture of the Poincar\'e section specific for integrable systems suggests that the full coset-space analysis, that takes into account R-R fields could indeed give Lax formulation of the model.

There is yet another argument for such an energy transfer to also happen in integrable systems, that we provide in the following subsection.

\subsection{Family of Type IIB backgrounds}

Explicit form of the solution analysed in the previous subsection suggests that performing a set of T- and S-dualities one is able to convert all R-R fields \eqref{eq:ads4xcp3def_RR} into the fields of the NS-NS sector. Although, this will not be enough to stabilize the string at constant $z$ due to the non-vanishing dilaton, we will have an integrable (for $\g=0$) model with the same energy transfer as before. Hence, we perform a T-duality along the coordinate $x^1$ and then S-duality to get the following Type IIB background
\begin{equation}
    \begin{aligned}
        ds^2& = \fr{R^3}{4(z^3 - \g x^2)}\left(-(dx^0)^2+ (dx^2)^2 \right)+\frac{R^3}{4 z^3}dz^2 +\frac{4  z}{R^3}(dx^1)^2 + \fr{R^3}{z}ds^2_{\CP^3}, \\
        e^{2\f} & = \fr{z}{z^3 - \g x^2},\\
        B&= -\frac{R^3}{4 (z^3 - \g x^2)} dx^0 \wedge dz +
        dx^1\wedge \big(c_{2\x} d\phi_1 +c_{\q_1} c_\x^2d\phi_2 + c_{\q_2} s_\x^2d\phi_3\big),
    \end{aligned}
\end{equation}
where we have used the same shorthand notations for cos and sin, and the metric $ds_{\CP^3}^2$ on $\CP^3$ is used in the parametrization \eqref{eq:cp3two}. Although the $x^1$ direction, along which the T-duality has been performed is not a cyclic, it is an isometry and formally Buscher rules are applicable. At the classical level of a point-like (in the AdS directions) string this is sufficient for our analysis. Alternatively, one may think of this background as of a family of Type IIB solutions with no reference to T-duality.

Now we see, that turning to the light-cone coordinates in the directions (0,1) leads to components $g_{++}$ and $g_{--}$, that were absent before. However, due to the fact that $b_{+\f_I} = \pm b_{- \f_I}$, where $I=1,2,3$ and the sign depends on how $x^1$ is defined in terms of $x^+$ and $x^-$, this does not change the overall approach. The Hamiltonian becomes $\mH= \mH_x$ (no factor of 2), that explicitly reads
\begin{equation}
    \begin{gathered}
        \mH = \frac{1}{4 z}\Big(\lambda _2^2 c_\x^2 \left(1-c^2_{\theta_1} c^2_\xi\right)-\lambda _3^2 c^2_{\theta _2}s^4_\xi 
        - 2 \lambda_2 \lambda_3 c_{\theta_1}c_{\theta _2} s^2_\xi c^2_\xi + \lambda _1 s^2_{2 \xi} \left(\lambda_2 c_{\theta_1}-\lambda_3 c_{\theta _2}\right) + \lambda_1^2 s^2_{2 \xi} + 
        \\
        + \lambda_3^2 s^2_{\xi} - 16 \gamma  p_2^2 x_2 z+16 p_2^2 z^4+16 p_3^2 z^4 + 16 z^2 s^{-2}_\xi p_{\theta _2}^2+16 z^2 c^{-2}_\xi p_{\theta _1}^2+4 z^2 p_{\xi }^2\Big).
    \end{gathered}
\end{equation}
The most important observation here is that the deformation parameter enters only via the combination $\g p_2 x^2$, and this is the only term that contains $p_2$ and $x^2$. This implies, that dynamics is independent of $x^2$, meaning $p_2=0$ and the deformation parameter becomes irrelevant and what we analyse is a T- and S-dual of the integrable sigma-model. Now, the general form of the Hamiltonian does not depend on the deformation parameter (at $p_2=0$), however equations of motion do not allow to fix $z=$const unless the angles are also fixed to some values. This shows that in a clearly integrable system one still may have an energy transfer between sectors, which might not be a surprise in general.

A comment must be made on preservation of integrability under such a transformation. To motivate such a statement let us again refer to the results of \cite{Orlando:2019rjg}, where it has been shown that a constant O$(d,d)$ transformation does not change a Lax pair. The T-duality transformation we do here is precisely such type of a transformation, and the corresponding sigma model must be close to the exceptional string of \cite{Arvanitakis:2018hfn}. The subsequent S-duality, certainly, does not change integrability properties of the string.

To see more details it is interesting to return to the example of the previous subsection where winding numbers $\l_1=1$, $\l_2=0$, $\l_3=2$ have been chosen. Here again $p_{\q_1}=const$ and the same integral of motion appears
\begin{equation}
    p_\x^2 + \fr{4 }{\cos^2 \x} p_{\q_1}^2 = 4\a.
\end{equation}
However now, the Hamiltonian extremely simplifies and reads
\begin{equation}
    \mH= 4 z^3 p_z^2 + 4\a z.
\end{equation}
From the equation we derive the second integral of motion
\begin{equation}
    z^3 p_z^2 - \a z = \b = \const,
\end{equation}
that allows to explicitly integrate equations for $z$. The result is
\begin{equation}
    z(\t) = \fr{\b}{\b^2(4 \t+\k)^2 -\a},
\end{equation}
where $\k$ is the integration constant. We see the same behaviour as the one obtained numerically: starting from any non-vanishing value of $z$ the string will be dragged to the horizon at $z=0$.

\subsection{Non-abelian \texorpdfstring{$D\wedge P\wedge P$}{DPP} deformations}

Another known non-abelian tri-vector deformation of $\AdS_4$ dubbed DPP in \cite{Bakhmatov:2020kul} is generated by dilation and three momentum generators. The corresponding tri-vector has the following form
\begin{equation}
    \W = \e^{abc}\r_a x^d \dt_d\wedge \dt_b\wedge \dt_c,
\end{equation}
where as before $a=0,1,2$. This deformation is two-parametric since a quadratic constraint has to be satisfied
\begin{equation}
    \r_0^2 - \r_1^2 - \r_2^2=0.
\end{equation}
The resulting background takes the following form 
\begin{equation}
    \begin{aligned}
        ds^{2} &= \frac{R^{2}}{4}\bigg(z^{3} + \rho_{a}x^{a}\bigg)^{-2/3}\bigg[ - (dx^{0})^{2} + (dx^{1})^{2} + (dx^{2})^{2} + \bigg(1 + \frac{\rho_{a}x^{a}}{z^{3}}\bigg)dz^{2} - \frac{1}{z^{2}}\rho_{a}dx^{a}dz \bigg] +
        \\
        & \quad+ \frac{R^{2}}{z}\bigg(z^{3} + \rho_{a}x^{a} \bigg)^{1/3}d\Omega^{2}_{7},
        \\
        F_{(4)}& = -\frac{3R^{3}z^{2}}{8}\bigg(z^{3} + \rho_{a}x^{a} \bigg)^{-2}dx^{0}\wedge dx^{1}\wedge dx^{2}\wedge dz.
    \end{aligned}
\end{equation}
Note that, in addition to explicit dependence on $x^0$ the background has mixed components $dx^0 dz$, that make gauge fixing following the approach of Section \ref{sec:basics} impossible. Moreover, one cannot simply choose $\r_0=0$ such as to remove time dependence because of the quadratic constraint. 

Instead we use the fact that the numerical approach we are following here is essentially local and notice that the DPP deformed metric by the following coordinate transformation
\begin{equation}
    x'{}^a = x^a + \fr{1}{2z}\h^{ab}\r_b
\end{equation}
can be mapped to that of the PPM, if the quadratic constraint $\h^{ab}\r_a\r_b=0$ is satisfied, where $\h^{ab} = \mathrm{diag}[-1,1,1]$. Indeed, in this case $\r_a x'{}^a = \r_a x^a$ and the overall prefactor does not change. On the other other hand the off-diagonal term $\r_a dx^a dz$ in the metric disappears.

Hence, one may think, that a subset of all PPM deformed backgrounds of AdS$_4\times \SS^7$ with the parameters satisfying $\h^{ab}\r_a \r_b=0$ are (at least locally) equivalent to DPP deformations. However, given the quadratic constraint, one is still not able to choose $\r_0=0$, and a dependence on $x^0$ remains. Nevertheless one is able to choose light-cone coordinates as before
\begin{equation}
    \begin{aligned}
        x_\pm = \fr{1}{\sqrt{2}}(x^0 \pm x^1),
    \end{aligned}
\end{equation}
and gauge fix $x^- = \t$. Then the Hamiltonian does not depend on the time coordinate $\t$, and is in fact precisely the same as in the PPM case with $\r_0=0, \r_1=\g, \r_2=0$ with replacement $x^1 \to x^+$. Therefore, in this case the whole dynamics does not change and Poincar\'e section are completely the same.

\section{Conclusions}
\label{sec:conclusions}

In this work we investigate classical integrability of the closed string on particular tri-vector deformed backgrounds understood as a two-dimensional sigma-model. Together with the Type IIB string on the bi-vector deformed $\AdS_5\times \SS^5$ whose integrability has been established analytically, we consider Type IIA string on $\AdS_4\times \mathbb{CP}^3$ deformed by the abelian $\rmU(1)^3$ tri-vector deformation of \cite{Lunin:2005jy} and by the PPM and DPP tri-vector deformations of \cite{Bakhmatov:2020kul}. In the first case the deformation is along three commuting isometries of the projective plane, while in the second and third cases the isometries are from the AdS symmetry algebra and do not commute. While classical integrability of the underformed Type IIA superstring  on $\AdS_4\times \CP^3$ has been established in \cite{Arutyunov:2008if,Stefanski:2008ik,Sorokin:2010wn} analytically, no analytical results are available so far for its tri-vector deformations. Ultimately, one is interested in finding Lax pair for the system similarly to the results of \cite{Delduc:2013fga,Delduc:2013qra}, which appears to be a subtle problem given the absence of a sigma--model formulation of tri-vector deformations. As a first step towards the full picture we consider a numerical analysis of phase trajectories of the string on deformed backgrounds.

Signatures of integrability we are looking for are regular closed phase curves depicted as closed curves in a given Poincar\'e section, and Lyapunov exponents, that determine mutual divergence of any given two trajectories with time. The former depict section of invariant tori wrapped by phase trajectories. Since our analysis is numerical it requires assumptions about initial condition and embedding of the string in the background space-time. The most important assumption is that oscillatory modes have been completely truncated and the string is assumed to evolve as a rigid rod. However, to keep at least some track of the internal geometry we assume that the string wraps some of the internal cycles. The initial conditions explicitly listed for each case have been chosen such that the resulting pictures become most representative. One obtains similar pictures for other choices of initial conditions that are squeezed and/or have less points etc.

Equations of motion are derived from a gauge fixed Hamiltonian in the light-cone gauge, that includes Virasoro constraints. We solve these equations numerically obtaining full phase trajectory and depict its intersections with a particular plane. These Poincar\'e section are presented as figures in the main text. Such obtained Poincar\'e sections exhibit the typical for integrable systems picture of closed phase curves, that belong to sets of tori dissected by the plane. We show that while the tori slowly change their shape, the overall picture preserves for very large values of the deformation parameter. Lyapunov exponent, that we plot only for large values of the deformation parameter, drops very quickly, showing that two trajectories tend to converge, that is yet another characteristic feature of an integrable system.

A few comments concerning choice of the embedding ansatz seem to be appropriate. For numerical analysis to be applied one had to choose particular values for all variables, including those that determine the embedding (winding numbers, initial position and momentum). However, the choice we made is far from being a fine tuning and is rather a random choice of parameters such that the Wolfram Mathematica program handles the obtained equations well, in a descent time and draws pictures where curves corresponding to different initial values are well separated. In addition to the ones presented in the text here we have checked several other combinations of parameters all of which resulted in similar pictures, that is the reason for not including all the plots with minor differences.

Given the results presented here and their stability under change of parameters of the ansatz we conclude that there is a pretty good chance that the full dynamics of the Type IIA (super)string on the considered tri-vector deformed backgrounds is integrable in the Liouville sense. Inspired by the observations reported here one's next step would be to search for an analytical expression for the corresponding Lax connections. Especially one is interested in the non-abelian PPM tri-vector deformation as it is governed by the generalized Yang--Baxter equation (genCYBE) obtained in \cite{Malek:2019xrf,Sakatani:2019zrs} from the algebraic point of view and in \cite{Gubarev:2020ydf} in the context of supergravity. In contrast to the standard classical Yang--Baxter equation, whose solution can be used for example to construct Lax pair of a mechanical system, relation of its generalization to integrability is not known. Note however the review \cite{Gubarev:2023jtp} discussing possible collocations between genCYBE and Nambu systems, including M2-branes ending on M5-branes. We hope to report on the progress in this direction soon.

\section*{Acknowledgments}
The authors would like to thank Dmitri Bykov, Nikita Kolganov and Kirill Gubarev for helpful discussions. This work has been supported by Russian Science Foundation grant RSCF-20-72-10144.

\appendix


\section{Embedding of \texorpdfstring{$\CP^3$}{CP3} into \texorpdfstring{$\SS^7$}{S7}}

In this paper we are using two parametrizations of the 7-sphere whose reduction along the Hopf cycle gives $\CP^3$. One is based on coordinates adapted to charges of the corresponding ABJM fields and has been used in \cite{Lunin:2005jy}. The other gives the standard interval for the $\CP^3$ and the coordinate choice is symmetric under particular reflections. This embedding has been used for example in \cite{Bergman:2009zh}. In this Appendix we give more details on each.

Let us start with the first parametrization used in \cite{Lunin:2005jy}. Metric on the 7-sphere has the following form
The metric has the following form
\begin{equation}
    ds_{\SS^7}^2  =\sum_{i=1}^4\big(d\m_i^2 + \m_i^2 d\f_i^2\big),
\end{equation}
where the functions $\m_i$ are given by
\begin{equation}
    \begin{aligned}
        \m_1 & = \cos \q ,\\
        \m_2 & = \sin \q \cos \a, \\
        \m_3 & = \sin \q \sin \a \cos \b,\\
        \m_4 & = \sin \q \sin \a \sin \b.
    \end{aligned}
\end{equation}
In terms of such introduced angles the metric on the 7-sphere becomes
\begin{equation}
\label{eq:s7one}
    ds_{\SS^7}^2 = d\q^2 + s_\q^2 (d\a^2 + s_\a^2d\b^2) + c_\q^2 d\f_1^2 + s_\q^2 \big[ c_\a^2 d\f_2^2+ s_\a^2(c_\b^2 d\f_3^2 + s_\b^2 d\f_4^2\big].
\end{equation}
In terms of the functions $\m_i$ and the angles $\f_i$ four complex coordinates $X_i$ on $\RR^8$ defining the 7-sphere as $|X_1|^2 + \dots + |X_4|^2 = 1$ have the following simple form
\begin{equation}
    X_i = \m_i e^{i \f_i}.
\end{equation}
Now define new angles $\varphi_1,\varphi_2,\varphi_3$ as follows
\begin{equation}
    \begin{aligned}
        \f_1 & = \y+ \varphi_3, && \f_2  = \y- \varphi_3-\varphi_2, && \f_3  = \y+ \varphi_2 -\varphi_1 , && \f_4  = \y+ \varphi_1.
    \end{aligned}
\end{equation}
The new angles correspond to isometries acting on the coordinates $X_i$ (four fields of ABJM) as follows
\begin{equation}
    \begin{aligned}
        & \varphi_1: && ( X_1, X_2, e^{-i \varphi_1}X_3, e^{i\varphi_1}X_4), \\
        & \varphi_2: && ( X_1, e^{-i\varphi_2} X_2, e^{i \varphi_2}X_3, X_4), \\
        & \varphi_3: && ( e^{i \varphi_3}X_1, e^{-i\varphi_3}X_2, X_3, X_4).
    \end{aligned}
\end{equation}
The combination
\begin{equation}
    4\y = \f_1 + \f_2 + \f_3 + \f_4 
\end{equation}
parametrized the Hopf cycle. Metric on the $\CP^3$ is obtained by the standard Kaluza--Klein reduction along $\y$.

For the alternative parametrization let us parametrize (complex) coordinates of $\RR^8$ defining the 7-sphere as follows  
\begin{equation}
    \begin{aligned}
        X_1 & = \cos \x \cos \fr{\q_1}{2} e^{i \fr{\y_1 + \f_1}{2}},\\
        X_2 & = \cos \x \sin \fr{\q_1}{2} e^{i \fr{\y_1 - \f_1}{2}},\\
        X_3 & = \sin \x \cos \fr{\q_2}{2} e^{i \fr{\y_2 + \f_2}{2}},\\
        X_4 & = \sin \x \sin \fr{\q_2}{2} e^{i \fr{\y_2 - \f_2}{2}},
    \end{aligned}
\end{equation}
with the following ranges for the angles
\begin{equation}
    \begin{aligned}
        & 0 \leq \x \leq \fr \p2, && 0 \leq \y_i \leq 4\p, \\
        & 0 \leq \f_i \leq 2\p ,  && 0 \leq \q_i \leq \p.
    \end{aligned}
\end{equation}
These coordinates cover the whole $\RR^8$ once. The 7-sphere is defined again as $X_1^2 + \dots +X_4^2 =1$ and the metric reads
\begin{equation}
    \begin{aligned}
    ds_{\SS^7}^2 = d\x^2 & + \fr{1}{4}\cos^2 \x \big[(d\y_1 + \cos \q_1 d\f_1)^2 + d\q_1^2 + \sin^2 \q_1 d\f_1^2\big]\\
     & + \fr{1}{4}\sin^2 \x \big[(d\y_2 + \cos \q_2 d\f_2)^2 + d\q_2^2 + \sin^2 \q_2 d\f_2^2\big]
    \end{aligned}
\end{equation}
This length element can be represented in the form of a $\rmU(1)$ bundle over $\CP^3$
\begin{equation}
    ds_{\SS^7}^2 = ds_{\CP^3}^2 + (d\f + \w)^2,
\end{equation}
where $4 \f = \y_1 + \y_2$ and the 1-form is given by
\begin{equation}
    \w = \fr12 \big(\cos^2 \x - \sin^2\x\big) d\y + \fr12 \cos^2 \x \cos \q_1 d\f_1 + \fr12 \sin^2 \x \cos \q_2 d\f_2
\end{equation}
with $2 \y = \y_1 - \y_2$. The metric on the $\CP^3$ then takes the following form
\begin{equation}
\label{eq:cp3two}
    \begin{aligned}
    ds_{\CP^3}^2 & =  d\x^2  + \cos^2 \x  \sin^2 \x\bigg(d\y + \fr12 \cos \q_1 d\f_1 - \fr12 \cos \q_2 d\f_2\bigg)^2 \\
    &\quad + \fr{1}{4}\cos^2 \x \big(d\q_1^2 + \sin^2 \q_1 d\f_1^2\big)+\fr{1}{4}\sin^2 \x \big(d\q_2^2 + \sin^2 \q_2 d\f_2^2\big)
    \end{aligned}
\end{equation}

\bibliography{bib.bib}
\bibliographystyle{utphys.bst}
\end{document}